\documentclass{optica-article}

\journal{opticajournal} 

\articletype{Research Article}

\usepackage{lineno}
\linenumbers 
\DeclareMathOperator{\sinc}{sinc} 
\usepackage{siunitx}

\begin{document}
\nolinenumbers
\title{Two-dimensional Control of a Biphoton Joint Spectrum}

\author{Anatoly Shukhin,\authormark{1,*} Inbar Hurvitz,\authormark{2} Sivan Trajtenberg-Mills,\authormark{2} Ady Arie,\authormark{2} and Hagai Eisenberg \authormark{1}}

\address{\authormark{1}Racah Institute of Physics, Hebrew University of Jerusalem, Jerusalem 91904, Israel\\
\authormark{2}School of Electrical Engineering, Fleischman Faculty of Engineering, Tel Aviv University, Tel Aviv 69978, Israel}

\email{\authormark{*}anatoly.shukhin@mail.huji.ac.il} 


\begin{abstract*} 
Control over the joint spectral amplitude of a photon pair has proved highly desirable for many quantum applications, since it contains the spectral quantum correlations, and has crucial effects on the indistinguishability of photons, as well as promising emerging applications involving complex quantum functions and frequency encoding of qudits. Until today, this has been achieved by engineering a single degree of freedom, either by custom poling nonlinear crystal or by shaping the pump pulse. We present a combined approach where two degrees of freedom, the phase-matching function, and the pump spectrum, are controlled. This approach enables the two-dimensional control of the joint spectral amplitude, generating a variety of spectrally encoded quantum states - including frequency uncorrelated states, frequency-bin Bell states, and biphoton qudit states. In addition, the joint spectral amplitude is controlled by photon bunching and anti-bunching, reflecting the symmetry of the phase-matching function.

\end{abstract*}

\section{Introduction}
Superposition states of single photons and entangled states of photon pairs are the core of quantum optics and quantum information technologies. The ability to efficiently create such quantum states and manipulate their properties are among the main problems in the way of building long-distance quantum communications, quantum computing, and quantum sensors \cite{o2009photonic,flamini2018photonic}. Regarding state generation, one of the most promising approaches is to harness nonlinear optical effects, e.g., spontaneous parametric down-conversion (SPDC) \cite{eisaman2011invited,migdall2013single}. Indeed, for the last decades, SPDC has been serving as a source of high-purity heralded single-photons, entangled photon pairs, flying qubits, and other photonic states \cite{zhong2020quantum, couteau2018spontaneous}.

In order to build long-distance quantum communication links as well as scalable quantum networks, in many cases it is preferable to encode information in the photons' spectral degree of freedom (DOF)\cite{kues2019quantum,ponce2022unlocking,lu2018quantum} due to its advantages: i) a quantum state encoded in the frequency domain experiences virtually no decoherence in comparison with, for example, polarization-encoded quantum information, and thus information can be transferred through existing standard optical fiber networks \cite{lukens2017frequency}; ii) Unlike polarization, frequency encoding can carry multilevel quantum information (qudits) \cite{lu2019quantum}, which has benefits when comparing to qubits\cite{lu2022bayesian,wang2020qudits,cozzolino2019high,lima2006propagation,bulla2023nonlocal}. In order to take full advantage of the high-dimensional Hilbert space in the frequency domain, it is necessary to be able to tailor the spectral properties of the generated states.

For photon pairs generated via SPDC, this mainly implies controlling their joint spectral intensity (JSI). There are different ways to control the JSI. One of them is to use domain-engineered nonlinear crystals \cite{dosseva2016shaping}. Recently, it has been shown that shaping the JSI via this approach can produce high-purity heralded single photons \cite{baghdasaryan2023enhancing,dosseva2016shaping,graffitti2018independent,graffitti2017pure,pickston2021optimised,branczyk2011engineered}, create frequency-bin entanglement \cite{morrison2022frequency}, and perform entanglement swapping \cite{graffitti2020direct,merkouche2022heralding}. Another approach is active spectral/temporal shaping of the pump field using acousto-optic or electro-optic modulators, where generation of pure heralded single photons \cite{li2020generation}, tailoring the temporal-mode structure of the heralded photons \cite{ansari2018tomography,brecht2015photon}, Schmidt-mode selection \cite{brecht2014demonstration}, and time-domain quantum-optical-synthesis \cite{jin2022two} have been shown. Temporal/spectral shaping can also be applied to the photons after they are generated \cite{pe2005temporal}. All four frequency-bin Bell states have been synthesized via pump modulation along with the shaping of entangled photons \cite{PhysRevLett.129.230505}. The possibility of the 2D modulation of the JSI was theoretically shown based on spectrally-resolved two-photon interference \cite{li2023spectrally}. It was theoretically demonstrated that the combination of the domain-engineered crystals together with pump modulation allows for a tunable frequency-bin multi-mode squeezed states generation and can be a fruitful path toward discovering new quantum applications \cite{drago2022tunable}.

\begin{figure}
    \centering
    \includegraphics[width=0.7\columnwidth]{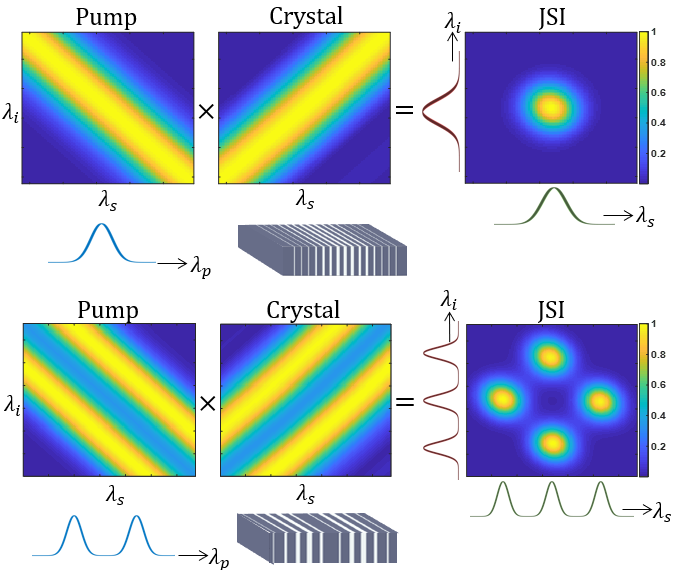}
    \caption{Illustration of the 2-dimensional control over the JSI. Top - Gaussian-shaped pump spectrum and PMF resulting a round JSI. Bottom - double-Gaussian-shaped pump spectrum and PMF resulting a 4 lobe JSI.}
    \label{fig:concept}
\end{figure}

In this work, we experimentally demonstrate for the first time that utilizing domain-engineered KTiOPO$_{4}$~(KTP) crystals along with modulation of a broadband femtosecond pulsed pump via a simple technique, using only passive optical elements, can generate a variety of two-photon and single-photon states with tailored spectral properties (Fig.\,\ref{fig:concept}). We also show how by using a single half-wave plate (HWP) or a linear polarizer after the nonlinear crystal, the JSI can be manipulated after the photons are generated, by photon bunching in the frequency domain. We demonstrate the generation of high purity frequency uncorrelated (separable) states, dual frequency correlated photon pairs as well as two of the four frequency-bin Bell states, frequency entangled biphoton 5-mode qudit states and biphoton frequency qutrit states, based on superposition of 4 pairs of signal/idler modes. The flexibility of generating different quantum states using the same setup is highly desired for quantum information applications. 

\section{Theoretical background}\label{sec:theory}

SPDC is a nonlinear optical process in which one photon from a pump field is annihilated in exchange for two photons that are simultaneously created into fields that are typically called signal and idler \cite{migdall2013single}. The frequencies and wave vectors of the three interacting fields obey energy and momentum conservation laws: $\omega_{p}=\omega_{s}+\omega_{i}$, and $k_{p} = k_{s} + k_{i}$.
Here we focus on the collinear (pump, signal and idler propagate in the same direction) degenerate (signal and idler share the same frequency) type-II (signal and idler are orthogonally polarized) SPDC due to its efficiency and usefulness from the experimental point of view.
The state vector of the two-photon field generated via SPDC in a bulk crystal can be calculated by first-order perturbation theory \cite{keller1997theory}:
\begin{equation}\label{eq:PDC_State}
\begin{split}
    |\psi\rangle =& \sqrt{1-\eta^{2}}|0\rangle 
    + \eta\iint  d\omega_{s}d\omega_{i}A(\omega_{s},\omega_{i}) a^{\dagger}_h(\omega_{s}) a^{\dagger}_v(\omega_{i}) |0\rangle,
\end{split}
\end{equation}
where $|0\rangle$ is the vacuum state, and $a^{\dagger}_h(\omega_{s})$ and $a^{\dagger}_v(\omega_{i})$ represent the creation operators into the signal and idler modes, with horizontal and vertical polarizations, respectively. $A(\omega_{s},\omega_{i})$ is the joint spectral amplitude (JSA) of the two-photon field describing the amplitude of generating the signal and idler photons at the respective frequencies $\omega_{s}$ and $\omega_{i}$. The conversion efficiency $\eta$ is proportional to the crystal nonlinear susceptibility $\chi^{(2)}$ and the crystal length $L$. The square of the absolute value of the JSA is the JSI that describes the probability density of generating the signal and idler photons at frequencies $\omega_{s}$ and $\omega_{i}$.

The JSA is given by a product of the phase-matching function (PMF) $\phi(\Delta k(\omega_{s},\omega_{i}))$ and the spectral amplitude of the pump field $A_{p}(\omega_{s}+\omega_{i})$ (Fig.\,\ref{fig:concept}) \cite{kaneda2021generation}. The former, being a Fourier transform of the crystal's poling function \cite{kaneda2021generation}, in the general case of an arbitrary poling design \cite{shiloh2012spectral}, can be written as:
\begin{equation}
    \phi(\Delta k(\omega_{s},\omega_{i})) = \chi^{(2)}\int^{\infty}_{-\infty} d(z)e^{i \Delta k(\omega_{s},\omega_{i})z}dz,
\end{equation}
where $d(z)$ is the poling function describing the sign of $\chi^{(2)}$ as a function of the longitudinal coordinate $z$. $\Delta k(\omega_{s},\omega_{i})$ is the longitudinal wave-vector mismatch which in the case of a constant poling period is given by \cite{fiorentino2007spontaneous,hum2007quasi}:
\begin{equation}\label{eq:QPM}
    \Delta k(\omega_{s},\omega_{i}) = k_{p}(\omega_{s}+\omega_{i})-k_{s}(\omega_{s})-k_{i}(\omega_{i})\pm\frac{2\pi m}{\Lambda}, 
\end{equation}
where $k(\omega_{l})=\omega_{l}n(\omega_{l})/c$, $n(\omega_{l})$ is the refractive index of the crystal at a frequency $\omega_{l}$ $(l=p,s,i)$, $m$ is the order of quasi-phase-matching and $\Lambda$ is the poling period. Thus, the PMF is essentially determined by the nonlinearity profile and the phase mismatch \cite{kaneda2021generation}.

The nonlinearity profile can be characterized by the poling period and duty cycle. The latter is the ratio between the lengths of domains of opposite signs. 
The poling period affects what wavelengths are phase-matched under a given pump, while the duty cycle affects generation amplitude at the phase-matched wavelengths (50\% duty cycle for maximum amplitude and 0\% for zero amplitude).

In the simplest case of a crystal without poling or with a constant poling period and duty cycle, the PMF can be expressed in the following way:
\begin{equation}\label{eq:sinc}
    \phi(\Delta k(\omega_{s},\omega_{i})) = \int^{L}_{0}e^{i\Delta k(\omega_{s},\omega_{i}) z}dz=L\sinc\left(\frac{\Delta k(\omega_{s},\omega_{i}) L}{2}\right)\exp\left({i\frac{\Delta k(\omega_{s},\omega_{i}) L}{2}}\right).
\end{equation}

The poling period and duty cycle that vary along the crystal, locally determine phase-matched wavelengths and their generation amplitude. By using a domain-engineered nonlinear crystal, it is possible to obtain the required PMF and therefore spectral properties of the generated photon pairs along the difference-frequency axis on a JSI graph.

Obtaining the desired spectral shape of the PMF $\phi(\Delta k(\omega_{s},\omega_{i}))$, was previously done by numerical optimization \cite{dosseva2016shaping}. However, this shape can be obtained analytically: by taking the Fourier transform of the PMF, which we denote $u(z)$, we can design the nonlinear crystal as follows \cite{shiloh2012spectral,leshem2014experimental}:

\begin{equation}
\begin{split}
d_{j}(z) = \text{sign}\left[\cos\left(\frac{2\pi}{\Lambda}z+\varphi_{j}(z)\right)-\cos(\pi q_{j}(z))\right],
\end{split}
\end{equation}
where $q(z) = \arcsin(|u(z)|/\pi)$, and $\varphi$ is the phase of $u(z)$. In this work, we used three different crystal designs, $d_j(z)$, where $j=0,1,2$ is for a sinc-, a Gaussian-, and a first-order Hermite-Gaussian-shaped PMF, respectively (Table~\ref{fig:table}). 

\begin{table}
    \centering
    \caption{A detailed table of the periodically poled, Gaussian and Hermite-Gaussian designs, providing crystal lengths, $\phi(\Delta k(\omega_{s},\omega_{i}))$, and figures that show the phase-matching function vs. $\Delta k$ (which is given in Eq.\ref{eq:QPM}). For the Gaussian and Hermite-Gaussian crystals $\sigma \approx 3.7\cdot 10^{-4} [1/\mu m]$. For the Gaussian and HG$_{1}$ designs we ignore the finite length of the crystal, since the efficiency at the edges of the crystal is negligible.}
    \includegraphics[width=0.8\columnwidth]{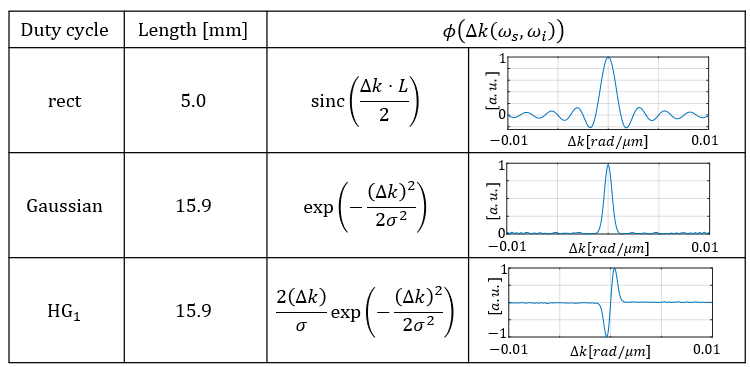}
    \label{fig:table}
\end{table}

The spectral amplitude of the pump, which is a bandwidth-limited sech$^{2}$ wave packet of duration $\tau_{p}$ (FWHM), can be approximated for simplicity by a Gaussian of the same FWHM as
\begin{equation}
A_{p}(\omega_{s}+\omega_{i}) = A_{0} \exp \left( {-\frac{(\omega_{p}-\omega_{p_{0}})^{2}}{2\sigma^{2}_{\omega}}} \right),
\end{equation}
where $A_{0}$ and $\omega_{p_{0}}$ are the amplitude and the central frequency of the pump field. The pump bandwidth (FWHM) is then defined as $\Delta\omega_{p}=2\sqrt{2\ln2}~\sigma_{\omega}=0.315\cdot2\pi/\tau_{p}$, where 0.315 is the time-bandwidth product for a sech$^{2}$ pulse.

The JSI reveals spectral properties, including frequency correlations of the two-photon field and therefore it is one of the main functions characterizing SPDC sources. Shaping the JSI enables one to generate quantum states ranging from pure heralded single-photons and entangled two-photon states \cite{u2003photon} to frequency-entangled qudits \cite{bernhard2013shaping,jin2016simple} and grid states \cite{fabre2020generation}.
As shown above, the JSI can be manipulated by modifying both the PMF and the pump spectrum.
In most cases, the PMF has a sinc-shaped profile due to the constant-period poling or homogeneous poling in the nonlinear crystals commonly used in experiments. On the other hand, the pump spectrum is usually the outcome of laser radiation, approximated by the Gaussian spectrum, as described earlier.
As a result, the spectrum of a generated state, as well as the state itself (since in some cases changing a spectrum leads to a different quantum state (see the following sections) are unlikely to have the required characteristics, such as purity and fidelity, approaching 1.

Nonetheless, as it will be shown in Secs.~\ref{sec:pump_mod}$-$\ref{sec:2d}, a simple way of pump modulation along with non-uniform crystal poling provides a powerful tool for building an adaptable source that is able to generate a variety of quantum states while keeping the changes in the experimental setup minimal.

\begin{figure}[t]
    \centering
    \includegraphics[width=\columnwidth]{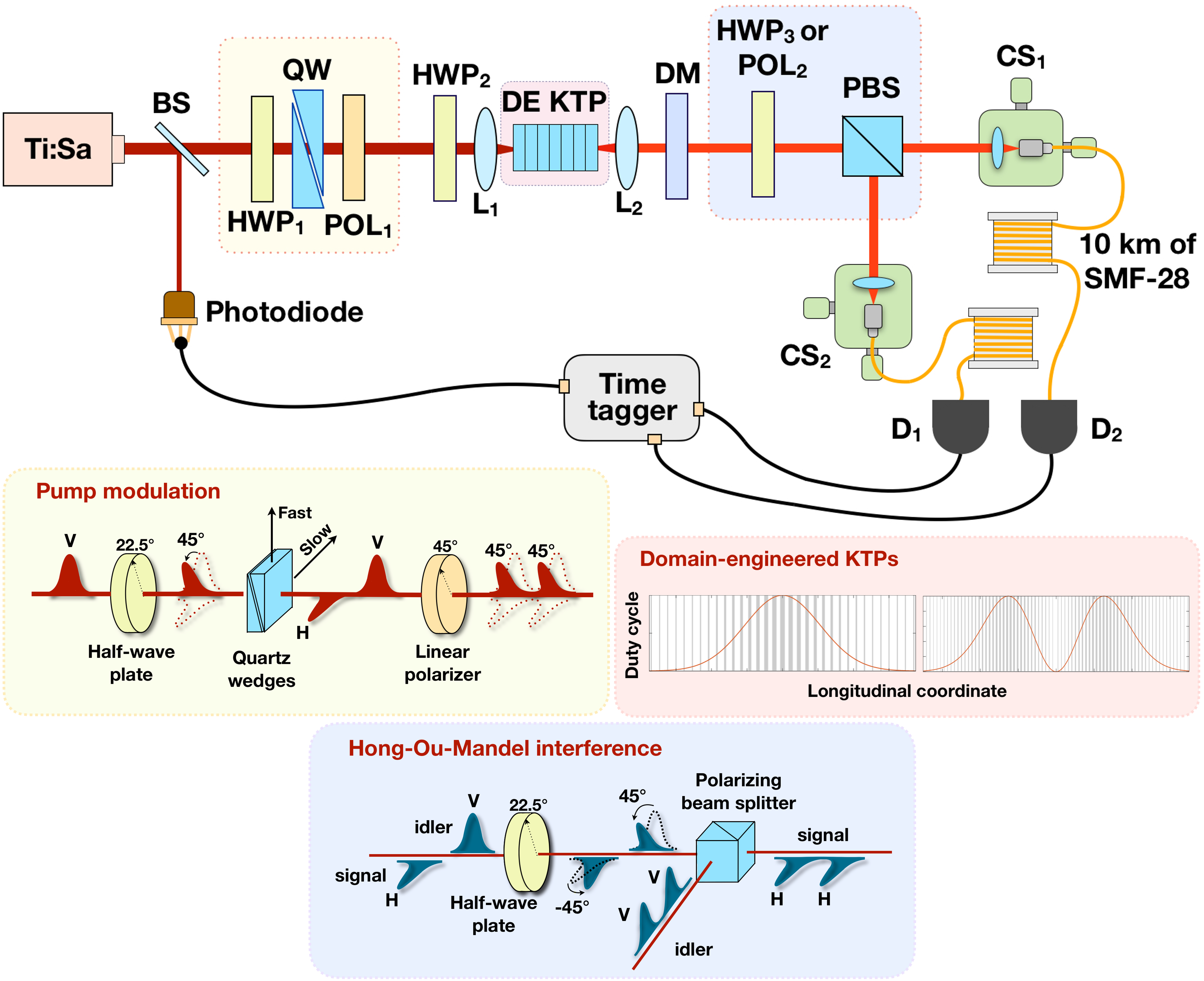}
    \caption{Experimental arrangement for observing SPDC in domain-engineered KTP crystals under the modulated pump. The radiation of the pulsed laser is split into two arms via a beam splitter (BS). The pump is modulated (shown in the yellow box), focused into the domain-engineered (DE) KTP crystal (shown in the pink box) where SPDC occurs, and then filtered out. Photon pairs, generated inside the crystal, are split into two arms by polarization, are coupled into single-mode fibers, and are detected by single-photon detectors. Arrival times of electrical signals from the detectors are compared using a time tagger. A small fraction of the pump power is detected by a photodiode, electrical signals from which are used as a reference clock allowing to measure the relative arrival time of the photon pairs. A HWP$_{3}$ or a POL$_{2}$ are used only for Hong-Ou-Mandel interference (shown in the purple box), described in sec.~\ref{sec:HOM}. See text for more details.}
    \label{fig:setup}
\end{figure}

\section{Setup description}

The experimental setup, illustrated in Fig.\,\ref{fig:setup}, involved the utilization of a tunable pulsed Titanium-Sapphire laser as the pump source with a pulse duration of $\approx$\,90\,fs, a pulse repetition rate of 76\,MHz, and a spectral width of approximately 7\,nm (FWHM). The central wavelength $\lambda_{p_{0}}$ was adjusted to 780\,nm.

The pump spectral modulation was acheived as follows (see the yellow box in Fig.\,\ref{fig:setup}): a vertically polarized pump pulse passes through a half-wave plate (HWP$_{1}$) rotated at $22.5^{\circ}$, followed by two movable quartz wedges (QW) with fast and slow axes in the \{H,V\} basis and a variable overall thickness in the range of $(2-18)\pm0.1$\,mm. The wedges split each diagonally polarized pump pulse into two pulses with orthogonal polarizations. The delay $\tau$ between them depends on the properties of the wedges: $\tau=|L_{\textrm{QW}}(n^{\textrm{gr}}_{\textrm{H}}-n^{\textrm{gr}}_{\textrm{V}})/c|$, where $L_{\textrm{QW}}$ is the overall thickness of the wedges, $n^{\textrm{gr}}_{\textrm{H}}$ and $n^{\textrm{gr}}_{\textrm{V}}$ are the group indices of quartz for the horizontally and vertically polarized light, respectively, and $c$ is the speed of light in vacuum. The mentioned thickness range allows a relative delay from $64$ to $572~(\pm2)$\,fs. Next, the two pulses are polarized using a linear polarizer (POL) positioned at an angle of $45^{\circ}$ relative to the reference frame of the wedges. The pump passes through another half-wave plate (HWP$_{2}$) to match its polarization to the KTP axes. Afterward, the pump is focused by the L$_{1}$ lens into the domain-engineered KTP crystal (Raicol Crystals Ltd.), where collinear type-II SPDC occurs.

After the crystal, the pump is filtered out using three dichroic mirrors (DM) with 40\,dB extinction for each mirror between the down-converted photons around 1560\,nm and the pump. The signal and idler photons are then collimated by the L$_{2}$ lens, split into two arms via a polarizing beam splitter (PBS), and coupled into single-mode fibers using coupling stages (CS$_{1}$, CS$_{2}$). The signal and idler photons are then detected by two InGaAs infrared single-photon avalanche detectors D$_{1}$, D$_{2}$ (Micro Photon Devices Srl), that are triggered by the laser pulses. The detection efficiency is $25\%$ at 1550\,nm, and the dark count rate in gated mode is $\sim0.8$\,kHz, at $6$\,V of excess bias voltage, gate duration of 1 ns, and \SI{30}{\micro\s} hold-off time. The detected electrical pulses arrival times were recorded using a time tagger (HRMTime, SensL) with a temporal resolution of 27\,ps. Based on the rates of detected coincidences ($0.7\pm0.1$\,kHz/mW) and single counts ($14\pm1$\,kHz/mW), and correcting for the hold-off time, as well as the losses in the setup and the detectors' efficiencies, the internal pair generation rate was $300\pm30$\,kHz/mW.

To measure the JSI of the generated photons, we used a time-of-flight spectroscopy \cite{avenhaus2009fiber,gerrits2011generation,zielnicki2018joint}. This technique leverages the natural capability of sufficiently long and dispersive fibers to perform a frequency-time Fourier transform on incoming light \cite{salem2013application}. The benefits of using this method are that it does not require alignment and offers a faster measurement process compared to alternative methods. We used 10\,km of SMF-28 optical fiber in both arms of the signal and idler photons. The fiber dispersion value is 18\,ps$/$nm$/$km at 1550\,nm, hence the spectral delay is 180\,ps/nm for a 10-km fiber. By comparing the arrival times of the correlated photons and the reference clock from the photodiode, we reconstruct the JSI.

The temporal resolution of our setup is limited by the total timing jitter, which consists of the jitters of the detectors, the time-tagger, and the clock photodiode. The combined effect of the latter two components was determined by analyzing the width of the distribution of the period between the times originating from the clock, which was found to be $\approx130$~ps. Quadrature summation of the mentioned jitter values results in a temporal resolution of 150\,ps (FWHM). 

In addition to timing jitter, the temporal resolution is also constrained by the marginal (unconditional) duration of the single photons being analyzed. However, in our case, the single photon duration is shorter than 1\,ps, which is significantly smaller than the timing jitter and has a negligible impact on the overall resolution.

Taking into account the dispersive fiber length tolerance (5\% of the fiber length),
the resolution in wavelength $\Delta\lambda$ was calculated to be $0.83\pm0.04$\,nm, using the relation $\Delta\lambda=\Delta t/DL_{\textrm{fiber}}$, where $L_{\textrm{fiber}}$ is the fiber length and $D$ is the fiber dispersion. The resolution can be further improved by using longer fibers and higher dispersion, usually at the expense of higher photon loss.

\section{Experimental results}
\subsection{Domain-engineered nonlinear crystals}

\begin{figure}[t!]
\centering
    \includegraphics[width=0.7\columnwidth]{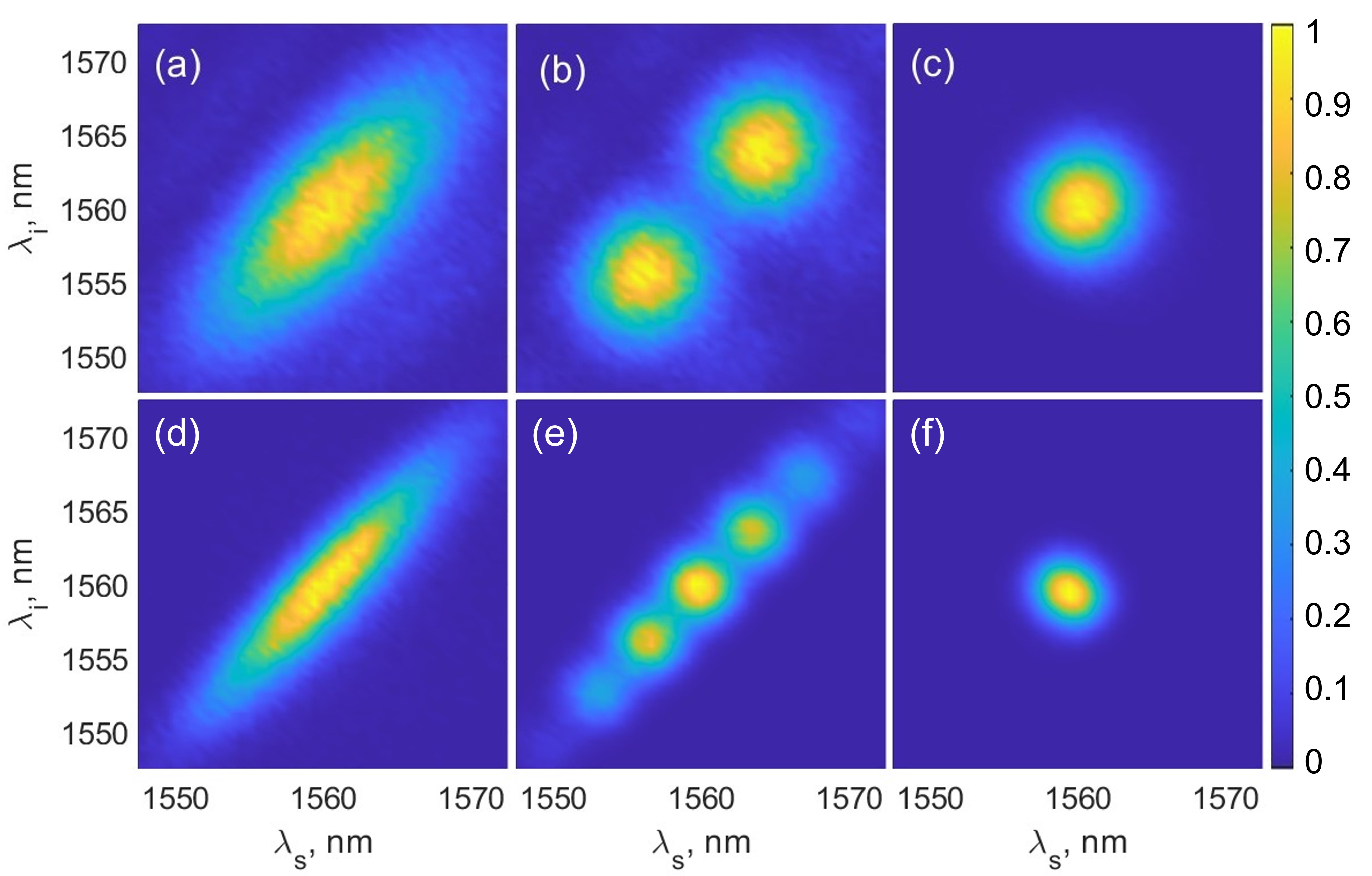}
\caption{JSI graphs, obtained from the 5\,mm long crystal (top row) and the 15.9\,mm long crystal with the single-lobe PMF (bottom row), (see Table\ref{fig:table} for details). (a) and (d) The non-modulated Gaussian pump of 7\,nm bandwidth. (b) and (e) The modulated pump; $\tau=226\pm2$\,fs, $L_\textrm{QW}=7.0\pm0.1~$mm and $\tau=382\pm2~$fs, $L_\textrm{QW}=12.0\pm0.1~$mm, respectively. (c) and (f) The non-modulated pump of 3\,nm and 1\,nm bandwidth, respectively. The graphs (a),(d),(c), and (f) are obtained without HWP$_{1}$, QW, and POL in the setup (see Fig.\,\ref{fig:setup}). Fig. (b) represents the frequency-bin $|\Phi^{-}\rangle$ Bell-state. The pump in (a) and (d) corresponds to the first column of Table\,\ref{table: pump}, while the pump in (b) and (e) corresponds to the second and third columns, respectively.}
\label{fig:pump_mod}
\end{figure}

Three KTP crystals, each having a constant poling period of $46\,\mu m$, were used in the experiments (see Table \ref{fig:table} for details). The first crystal is 5\,mm long and has a constant duty cycle of $50\%$. The second crystal has a duty cycle that gradually varies from the crystal's center (50\%) towards the edges (3\% due to the finite domain width), describing a Gaussian function (zeroth-order Hermite-Gaussian HG$_{0}$) which results in an apodized nonlinearity along the crystal \cite{phillips2013apodization,suchowski2014adiabatic,pickston2021optimised}. The first reason for apodization is to eliminate the side lobes that appear in an ordinary $\text{sinc}$-shaped phase-matching function. These side lobes, being arranged at an angle to the signal/idler wavelength axes (see, for instance, Fig.\,\ref{fig:pump_mod}a), reduce the purity of heralded single-photon states, generated in non-apodized crystals \cite{migdall2013single,baghdasaryan2023enhancing,dosseva2016shaping,graffitti2018independent,graffitti2017pure,pickston2021optimised,branczyk2011engineered}.

\begin{figure}[t!]
  \centering
    \includegraphics[width=\columnwidth]{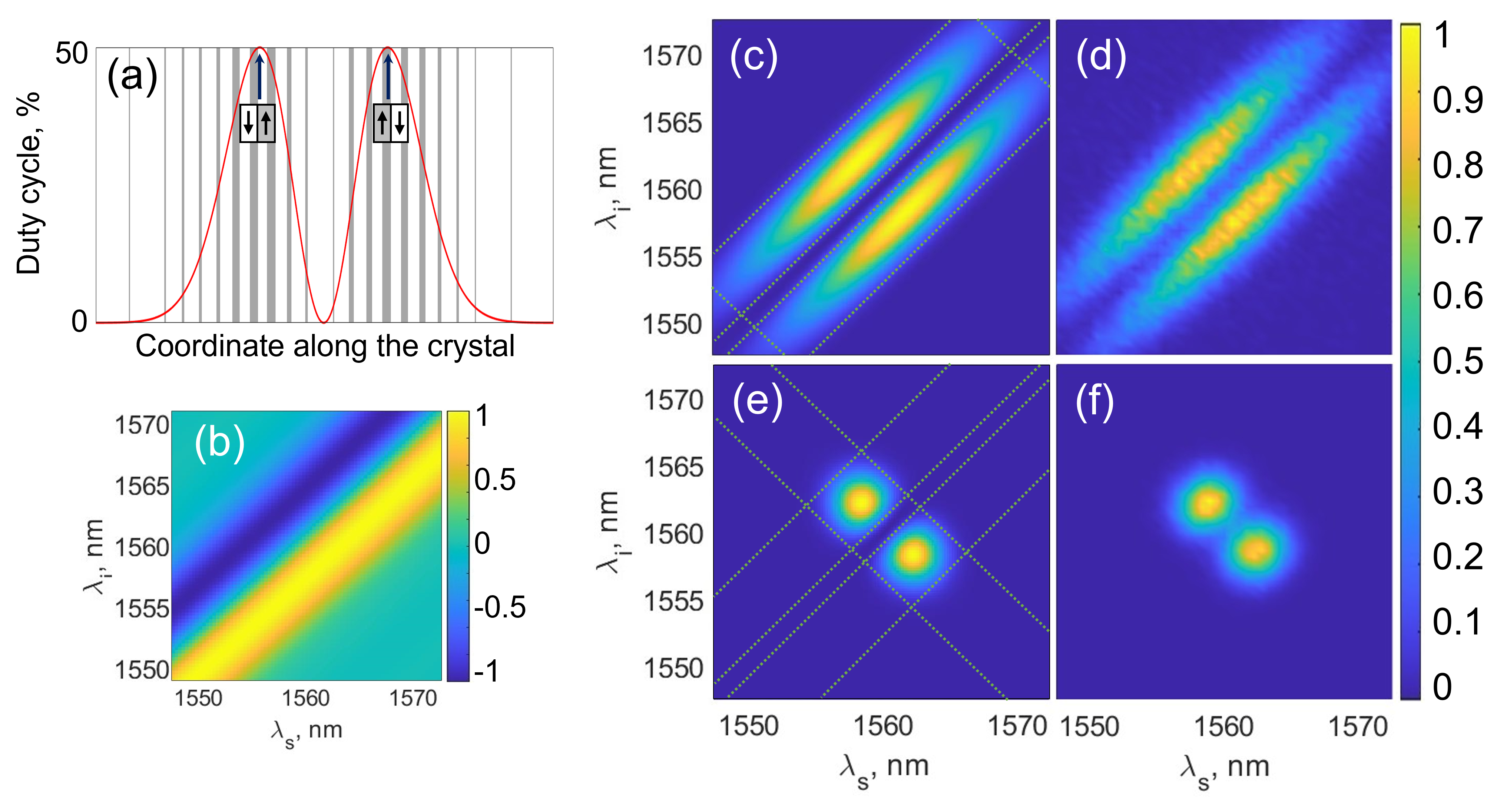}
    \caption{SPDC in the domain-engineered 15.9 mm long crystal, that is described by the double-lobe PMF (see Table\ref{fig:table}, bottom line). (a) poling design, where the domains in the two halves of the crystal are arranged in the opposite order (b) calculated PMF, (c) and (d) calculated and experimentally measured JSI without pump filtering, respectively, (e) and (f) calculated and experimentally measured JSI under the filtered pump of 1.5\,nm bandwidth, respectively. The generated state is the frequency-bin $|\Psi^{-}\rangle$ Bell state. The lines in (c) and (e) correspond to the $1/e^2$ of the PMF and pump intensity distributions.}
    \label{fig:15.9mm_2lobe} 
\end{figure}

The poling pattern of the third crystal (Fig.\,\ref{fig:15.9mm_2lobe}a) results in a first-order Hermite-Gaussian $\text{HG}_1$ PMF, consisting of two main lobes (having positive and negative nonlinear response) with no side-lobes (Fig.\,\ref{fig:15.9mm_2lobe}b). Multiplying this PMF by the pump spectrum and taking the absolute squared value, results with the JSI shown in Fig.\,\ref{fig:15.9mm_2lobe}c. The experimentally measured JSI is shown in Fig.\,\ref{fig:15.9mm_2lobe}d. An interesting feature of this JSI is that while each lobe, being diagonally elongated, corresponds to the frequency-correlated state, the two lobes as a whole are anti-diagonally arranged and demonstrate the frequency anti-correlated state. As opposed to the other methods that rely on interference of the generated biphotons (such as in Ref. \cite{jin2022two}), the resulting modulation from the domain-engineered crystals does not require precise stabilization and alignment of interfering beams, since the interference occurs inside the nonlinear crystal.

If the pump is filtered such that its bandwidth is equal to that of the single lobe of the PMF, the JSI results in two round lobes, positioned along the difference-frequency axis and equidistant from the exact degenerate regime. This corresponds to a frequency-bin $|\Psi^{-}\rangle$ Bell state \cite{PhysRevLett.129.230505}: $|\omega^{s}_{1}\omega^{i}_{2}\rangle-|\omega^{s}_{2}\omega^{i}_{1}\rangle$, where $\omega^{s}_{1,2}$ and $\omega^{i}_{1,2}$ are the central frequencies of each lobe for the signal and idler field, respectively. The sign between the two terms in the above expression reflects the relative phase between the two lobes of the PMF, which in this case is $\pi$. The simulated and measured JSIs for this case are shown in Fig.\,\ref{fig:15.9mm_2lobe}e and Fig.\,\ref{fig:15.9mm_2lobe}f, respectively. Notice, that all the graphs in this paper are normalized to a maximum value of 1. A convenient way to quantify the degree of entanglement of a given state is the cooperativity parameter $\mathcal{K}$ (also known as the Schmidt number) \cite{u2003photon}. In the case of the ideal Bell state, the cooperativity is equal to 2, whereas for the JSI in Fig.\,\ref{fig:15.9mm_2lobe}f it is reduced to $1.706\pm0.003$ due to the presence of background noise.

Using a crystal with two in-phase phase-matching lobes
allows for creation of the frequency-bin $|\Psi^{+}\rangle$ Bell as well.

Generally, the domain engineering technique gives the ability to tailor the generated state, affecting the generation probability amplitude within a nonlinear crystal. The following sections will discuss other options to control the JSI before and after the nonlinear crystal.

\subsection{Pump modulation}\label{sec:pump_mod}

\begin{table}
    \centering
    \caption{A detailed table of the modulation of the pump spectra. Three pump spectra used in the experiments are shown here in both the time and the frequency domains.}
    \includegraphics[width=0.8\columnwidth]{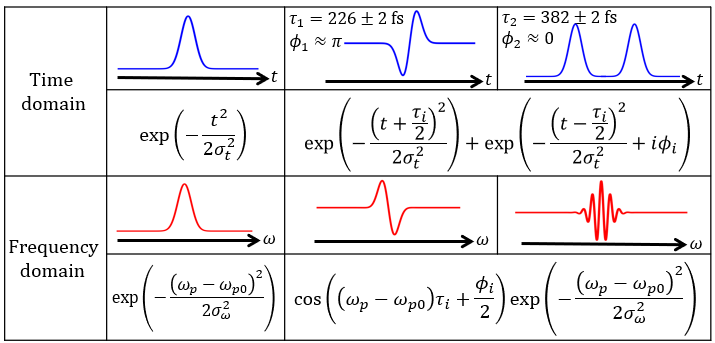}
    \label{table: pump}
\end{table}

Splitting each pump pulse into two pulses with a delay (determined by the thickness and refractive indices of the quartz wedges) allows for a modulation of the pump spectrum (Table \ref{table: pump}). This translates into a modulation of the JSI along the sum-frequency axis. Specifically, a Fourier transform (denoted below as $\mathcal{F}$) of the two Gaussian wave packets with a carrier frequency $\omega_{p_{0}}=2\pi c/\lambda_{p_{0}}$ and with a relative delay $\tau$ can be written as:
\begin{eqnarray}
    \mathcal{F}\left\{e^{i\omega_{p}t} \left( e^{\frac{-\left(t+\frac{\tau}{2}\right)^{2}}{2\sigma^{2}_{t}}}+e^{\frac{-\left(t-\frac{\tau}{2}\right)^{2}}{2\sigma^{2}_{t}}}e^{i\phi} \right)+\textrm{c.c.}\right\} =e^{\frac{-\left(\omega_{p}-\omega_{p_{0}}\right)^{2}}{2\sigma^{2}_{\omega}}}\cos\left((\omega_{p}-\omega_{p_{0}})\tau+\frac{\phi}{2}\right),
\end{eqnarray}
where $\phi$ is the phase difference between the two pulses accumulated during their propagation through the quartz wedges of thickness $L_\textrm{QW}$. By employing quartz wedges with variable thicknesses, we can adjust the pump spectrum modulation frequency and the position of its maxima and minima. The different modulations of the pump's temporal shape are shown in Table \ref{table: pump}, together with their spectral shape.

\subsubsection{5\,mm periodically poled KTP crystal}

The effect of pump modulation on the JSI for the 5\,mm long crystal can be seen in Fig.\,\ref{fig:pump_mod} (top row). Without modulation, the JSI is an ellipse elongated along the sum-frequency axis (Fig.\,\ref{fig:pump_mod}a). Such a state corresponds to the frequency-correlated photon pairs. Once the pump is modulated, the initial single-lobe ellipse turns into multiple lobes. Adjusting the delay $\tau$, it is possible to attain two round lobes (Fig.\,\ref{fig:pump_mod}b). The state in this case corresponds to the frequency-bin $|\Phi^{-}\rangle$ Bell-state \cite{PhysRevLett.129.230505}: $|\omega^{s}_{1}\omega^{i}_{1}\rangle-|\omega^{s}_{2}\omega^{i}_{2}\rangle$. The minus sign reflects the fact that the phase difference between the two adjacent lobes is $\pi$ since the modulating cosine function of the pump changes its sign on every zero crossing. $\omega_{1}$ and $\omega_{2}$ are the central frequencies of the two lobes, which are correlated for the signal and idler field. The cooperativity parameter in this case is $1.933\pm0.003$. It is also possible to use two pairs of quartz wedges of equal thicknesses along with a half-wave plate at $22.5^\circ$ placed in between, to obtain two in-phase lobes in the pump spectrum. Such spectrum generates the $|\Phi^{+}\rangle$ Bell state without the need for active electro-optical components.
Spectral filtering of the pump selects one of the lobes, resulting in a round-shaped JSI. In this case, the generated photon pairs are frequency-uncorrelated, and the heralded single photons are in a pure state \cite{migdall2013single}. The condition for a round-shaped JSI (assuming non-modulated pump and the single-lobe PMF) is the equality between the widths of the JSI along the pump and PMF directions. This is achievable by the pump filtering whenever the PMF's bandwidth is narrower than that of the pump. For the 5\,mm long crystal, this condition corresponds to a pump filter of 3\,nm bandwidth. The measured JSI under such filtering is demonstrated in Fig.\,\ref{fig:pump_mod}c. The Schmidt decomposition analysis \cite{law2000continuous} of this JSI resulted in heralded single-photon purity ($1/\mathcal{K}$) as high as $0.996\pm0.003$. This is based on the assumption that the shape of the JSI mirrors the shape of the JSA, disregarding any joint phase effects, such as $\exp(i
\omega_{s}\omega_{i}t^{2})$ in the case of a chirped pump, which render the JSA non-factorable in terms of phase.  Thus, it is possible to efficiently switch between the generation of frequency-correlated and frequency-uncorrelated states, which increases the versatility of this source.

\subsubsection{15.9\,mm single-lobe phase matched KTP crystal}
Using longer crystals leads to a narrower bandwidth of the PMF. Spectrally uncorrelated two-photon states may still be generated by pump filtering. Fig.\,\ref{fig:pump_mod}d and Fig.\,\ref{fig:pump_mod}f show the JSIs obtained from a 15.9\,mm long crystal without and with a spectral bandpass filter of 1.5\,nm bandwidth placed in the pump beam. As seen in the graphs, such a filter bandwidth matches the PMF's bandwidth, generating a round-shaped JSI (without side lobes due to apodization). The heralded single-photon purity is $0.988\pm0.003$. 

Frequency-bin entanglement between two photons does not necessarily equate to entanglement between two qubits. The JSI shown in Fig.\,\ref{fig:pump_mod}b indeed describes the frequency-entangled state of two qubits, since each field is in the superposition of two frequency bins (lobes): $\omega_{1}$ or $\omega_{2}$. Nevertheless, a longer delay $\tau$ increases the modulation frequency of the pump and hence also the number of lobes in the JSI (Fig.\,\ref{fig:pump_mod}e).
As a result, each biphoton is in a superposition of more than two spectral bins, leading to the generation of a frequency-entangled two-qudit state 
\begin{equation}
|\psi\rangle=\sum_{n} (-1)^{n} c_{n}|\omega^{s}_{n}\omega^{i}_{n}\rangle,
\end{equation}
where $n$ and $c_{n}$ are the indices and the weights of the frequency bins. The cooperativity parameter in this case is $2.745\pm0.003$. Notice that the described two-qubit and two-qudit frequency-entangled states are pure only if each lobe has a round shape and the projections of the lobes onto the frequency axes are not overlapping. By choosing the delay $\tau$ appropriately, the first requirement may be satisfied for different crystal designs and lengths, as well as for different pump bandwidths. The second requirement is satisfied by further controlling the shape of the pump pulse.
We should note again that using multiple quartz wedges with different thicknesses can enable more intricate modulation in the JSI, providing the ability to control the relative phase between different lobes in the pump and photon pairs' spectrum. Frequency-bin qudit states were recently reported in \cite{morrison2022frequency}, by using structured crystals. In contrary to what is shown here, in that case the states are aligned along the orthogonal line of the joint spectrum, with different frequencies of the signal and idler in each of the separate lobes.

\subsection{Two-dimensional control of the JSI}\label{sec:2d}
\begin{figure}[t!]
    \centering
    \includegraphics[width=0.7\columnwidth]{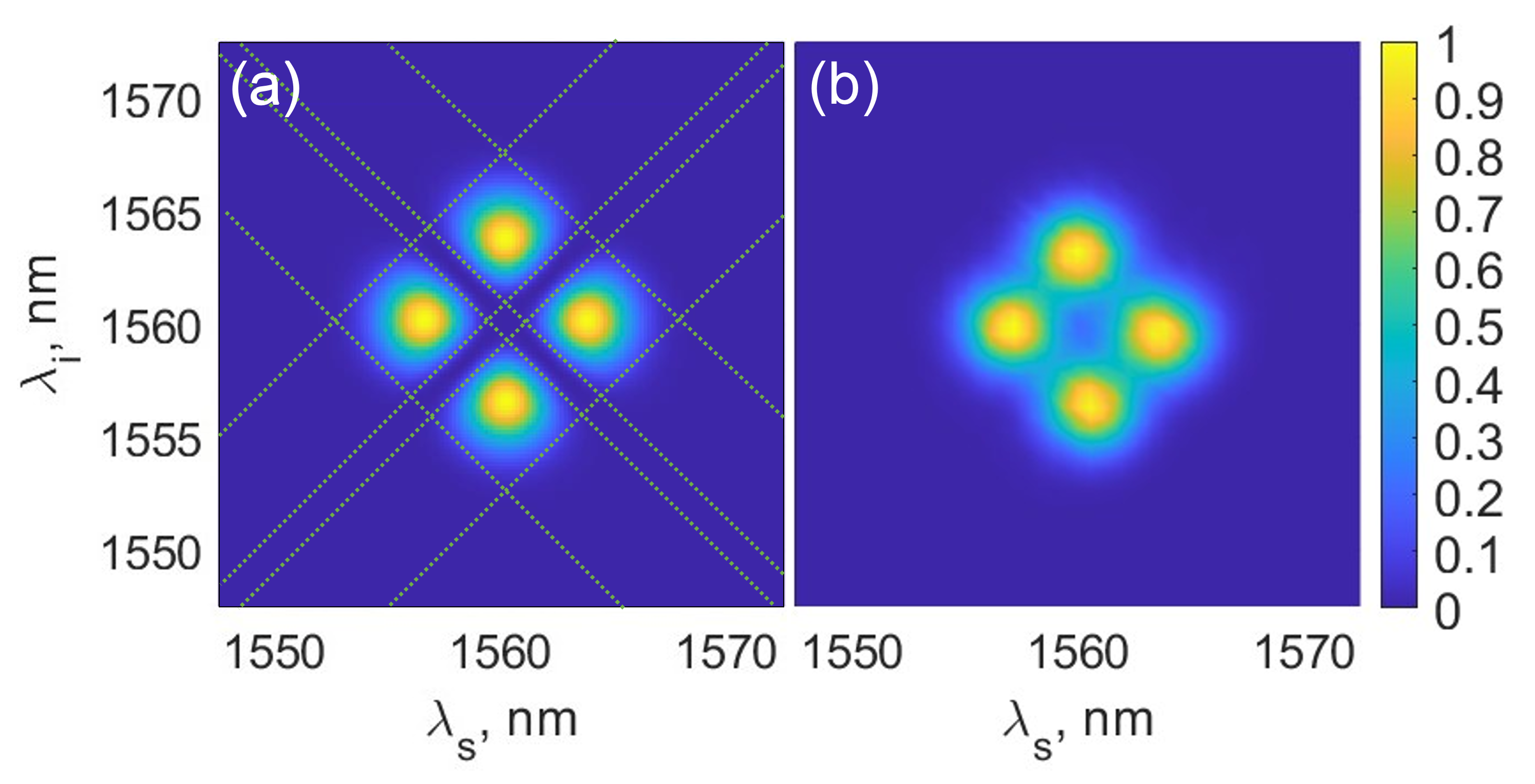}
    \caption{A JSI from the 15.9\,mm long crystal with the double-lobe PMF (see Table\ref{fig:table} bottom line) and modulated pump spectrum: (a) simulated, (b) experimentally measured. The lines in (a) correspond to the $1/e^2$ of the PMF and pump intensity distributions.}
    \label{fig:4lobes}
\end{figure}

The combination of the modulated pump and the domain-engineered nonlinear crystal allows for two-dimensional control of the JSI. The control directions are along the sum- and difference-frequency axes. The 15.9\,mm long KTP crystal, characterized by the double-lobe PMF, enables the generation of unusual quantum states such as a frequency-entangled two-qutrit state. The calculated and measured JSIs, consisting of four lobes are presented in Fig.\,\ref{fig:4lobes}a and Fig.\,\ref{fig:4lobes}b, respectively.

In order to create this JSI, the pump spectrum was modulated so that it consists of two lobes, whose bandwidths are equal to that of the PMF lobes. Under this condition, the four lobes are round. In the basis of the marginal spectra, each photon is projected onto three spectral lobes that correspond to the basis states of a qutrit. The generated two-qutrit state is neither correlated nor anti-correlated, in contrast to the regular frequency-entangled states. This two-photon state can be written as
\begin{equation}
|\psi\rangle=|\omega^{s}_{2}\rangle\otimes\left(|\omega^{i}_{1}\rangle+|\omega^{i}_{3}\rangle\right)-\left(|\omega^{s}_{1}\rangle+|\omega^{s}_{3}\rangle\right)\otimes|\omega^{i}_{2}\rangle,
\end{equation}
where the lower indices $1,2$, and $3$ correspond to the left, central, and the right maxima of each field, respectively. The state, however, can be turned into either correlated or anti-correlated by post-selecting corresponding pairs of lobes. The cooperativity parameter for the calculated JSI (Fig.\,\ref{fig:4lobes}a) is 2, corresponding to the two-mode entanglement: \includegraphics[scale=0.1]{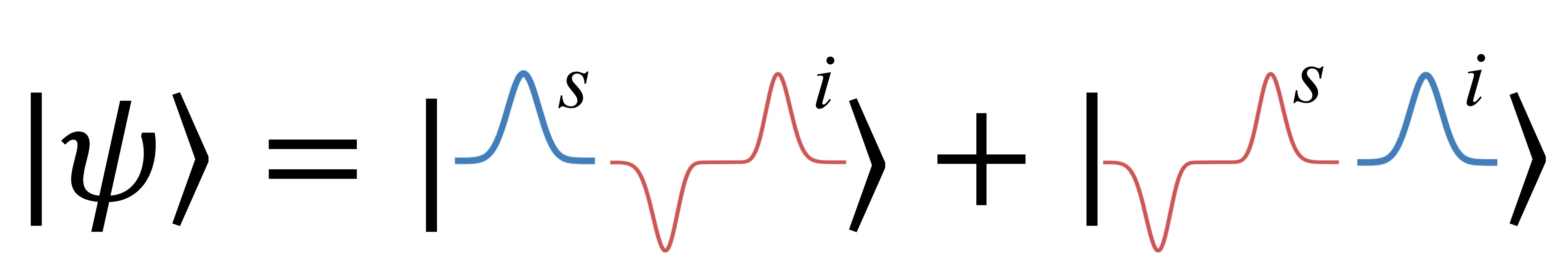}. The same parameter for the measured JSI (Fig.\,\ref{fig:4lobes}b) is reduced to $1.506\pm0.003$ due to the presence of background noise. More recently, this state has been theoretically investigated in the context of the time-frequency quantum metrology \cite{descamps2023quantum}.

An interesting aspect of this state is revealed when the signal and idler photons are sent to Alice and Bob, respectively. Whenever Alice detects her photons in the state  $|\omega_{2}\rangle$, she gets no information about the corresponding outcome of Bob, which can be either $|\omega_{1}\rangle$ or $|\omega_{3}\rangle$. On the other hand, whichever Bob's result in this case is, he knows unambiguously what is the outcome of Alice. Conversely, taking into account the symmetry of the state, if Alice measures either $|\omega_{1}\rangle$ or $|\omega_{3}\rangle$, the roles are switched.

Increasing the number of peaks in the pump and/or in the PMF allows for the generation of frequency-bin cluster states \cite{reimer2019high,hurvitz2023frequency} and paves the way for the generation of more intricate states, such as Gottesman, Kitaev and Preskill (GKP) states, that can be used for quantum error correction \cite{fabre2020generation}.

\subsection{Shaping the JSI via Hong-Ou-Mandel interference}\label{sec:HOM}
Generation of states with multiple lobes along the sum-frequency direction, such as those presented in Fig.\,\ref{fig:15.9mm_2lobe} and Fig.\,\ref{fig:4lobes} is possible even if the crystal is not domain-engineered. In this method, after the photons are already created, the JSA interferes with its own spectrally-flipped (around $\omega_{s}=\omega_{i}$ axis) replica. This interference can be achieved in two ways. The first method is to rotate the polarization of the signal and idler photons by $45^{\circ}$ by placing a half-wave plate (HWP$_{3}$ in Fig.\,\ref{fig:setup}) at $22.5^{\circ}$. The biphoton state from the source is described by the Fock representation in Eq.\,\ref{eq:PDC_State}. After the wave-plate rotation, the state is transformed into \cite{wang2006quantum}
\begin{equation}
\begin{split}\label{eq:JSA_HOM}
    |\psi\rangle=&\frac{1}{2}\iint d\omega_1 d\omega_2 A(\omega_1,\omega_2)  \times 
    \\&[a^\dagger_h (\omega_1) a^\dagger_h (\omega_2)+a^\dagger_h (\omega_1) a^\dagger_v (\omega_2)-a^\dagger_v (\omega_1) a^\dagger_h (\omega_2)+
    a^\dagger_v (\omega_1) a^\dagger_v (\omega_2)] |0\rangle.
\end{split}
\end{equation}
At the following PBS (see Fig.\,\ref{fig:setup}), the first and last terms do not yield to eventual coincidences, while the two middle terms do. By re-labeling the second term that is left after the post-selection, the detected normalized state is
\begin{equation}
\begin{split}\label{eq:JSA_PS}
    |\psi\rangle=&\iint d\omega_1 d\omega_2 [A(\omega_1,\omega_2)-A(\omega_2,\omega_1)]a^\dagger_h (\omega_1) a^\dagger_v (\omega_2)|0\rangle.
\end{split}
\end{equation}
As long as the JSA is symmetric (even) for the exchange between $\omega_1\Leftrightarrow\omega_2$, $A(\omega_1,\omega_2)=A(\omega_2,\omega_1)$ and these two terms cancel each other, a manifestation of the Hong-Ou-Mandel bunching effect. In the case of an anti-symmetric (odd) JSA $A(\omega_1,\omega_2)=-A(\omega_2,\omega_1)$ these two terms interfere constructively and the photons anti-bunch. The polarization walk-off $\tau_{wo}$ between the two photons in the generating crystal at the reference frame of the $\omega_1$ photon, can also break the symmetry of the JSA, which becomes $A(\omega_1,\omega_2)e^{-i\omega_2\tau_{wo}}$. Substituting a JSA from a symmetric PMF (such as from the $\text{HG}_0$ crystal) and walk-off into Eq.\,\ref{eq:JSA_PS}, the observed JSI is \cite{kuo2016spectral,jin2015spectrally,chen2021temporal,wang2006quantum,chen2023spectrally,jin2015spectrally,gerrits2015spectral,orre2019interference,jin2016simple,chen2021temporal}
\begin{equation}\label{eq:JSI_HWP}
\begin{split}
    I_{\text{HWP}}=&|A(\omega_1,\omega_2)e^{-i\omega_2\tau_{wo}}-A(\omega_2,\omega_1)e^{-i\omega_1\tau_{wo}}|^2
    =|A(\omega_1,\omega_2)|^2\sin^2 \left(\frac{\omega_2-\omega_1}{2}\tau_{wo}\right).
\end{split}
\end{equation}
The biphotons of the $\omega_1=\omega_2$ degenerate line are indistinguishable, and thus result in no coincidences. The JSI oscillates along the sum-frequency direction. If the JSA without walk-off is anti-symmetric (such as from the $\text{HG}_1$ crystal) the sine function in the JSI is replaced by a cosine function.

A second method for interfering the biphoton spectrum with itself is by placing a linear polarizer (POL$_{2}$ in Fig.\,\ref{fig:setup}) at $45^{\circ}$ before the PBS. The original state of Eq.\,\ref{eq:PDC_State} can be written in the $\pm45^{\circ}$ diagonal polarization basis as
\begin{equation}
\begin{split}\label{eq:JSA_POL}
    |\psi\rangle=&\frac{1}{2}\iint d\omega_1 d\omega_2 A(\omega_1,\omega_2) \times
    \\&[a^\dagger_p (\omega_1) a^\dagger_p (\omega_2)+a^\dagger_p (\omega_1) a^\dagger_m (\omega_2)-a^\dagger_m (\omega_1) a^\dagger_p (\omega_2)+
    a^\dagger_m (\omega_1) a^\dagger_m (\omega_2)] |0\rangle,
\end{split}
\end{equation}
where $a^\dagger_p$ and $a^\dagger_m$ represent the creation operators of a $+45^{\circ}$ and a $-45^{\circ}$ polarized photons, respectively. The polarizer transmits only the $p$-polarized photons, where only the first term can contribute two photons for the eventual coincidence measurement. As the PBS projects on the \{H,V\} basis, the normalized transmitted state in this basis is
\begin{equation}
\begin{split}\label{eq:JSA_POL_PBS}
    |\psi\rangle=&\frac{1}{2}\iint d\omega_1 d\omega_2 A(\omega_1,\omega_2) \times
    \\&[a^\dagger_h (\omega_1) a^\dagger_h (\omega_2)+a^\dagger_h (\omega_1) a^\dagger_v (\omega_2)+a^\dagger_v (\omega_1) a^\dagger_h (\omega_2)+
    a^\dagger_v (\omega_1) a^\dagger_v (\omega_2)] |0\rangle.
\end{split}
\end{equation}
Again, only the two middle terms can result in coincidence detection. Following a similar derivation as for the half-wave plate case, the opposite sign of the third term in Eq.\,\ref{eq:JSA_POL_PBS} results in the following JSI
\begin{equation}\label{eq:JSI_POL}
\begin{split}
    I_{\text{POL}}=&|A(\omega_1,\omega_2)e^{-i\omega_2\tau_{wo}}+
    A(\omega_2,\omega_1)e^{-i\omega_1\tau_{wo}}|^2
    =|A(\omega_1,\omega_2)|^2\cos^2 \left(\frac{\omega_2-\omega_1}{2}\tau_{wo}\right).
\end{split}
\end{equation}
As the polarizer erases the polarization difference between the two photons, in this case the biphoton spectrum and its flipped version interfere constructively along the degenerate line where the photons anti-bunch. The oscillation is along the perpendicular direction as in the previous case. The PMF, in this case, is symmetric (e.g. $\text{HG}_0$), but an anti-symmetric function will result in bunching along the degenerate line, and the cosine will be replaced by a sine function.

\begin{figure}[t!]
     \centering
         \includegraphics[width=0.7\columnwidth]{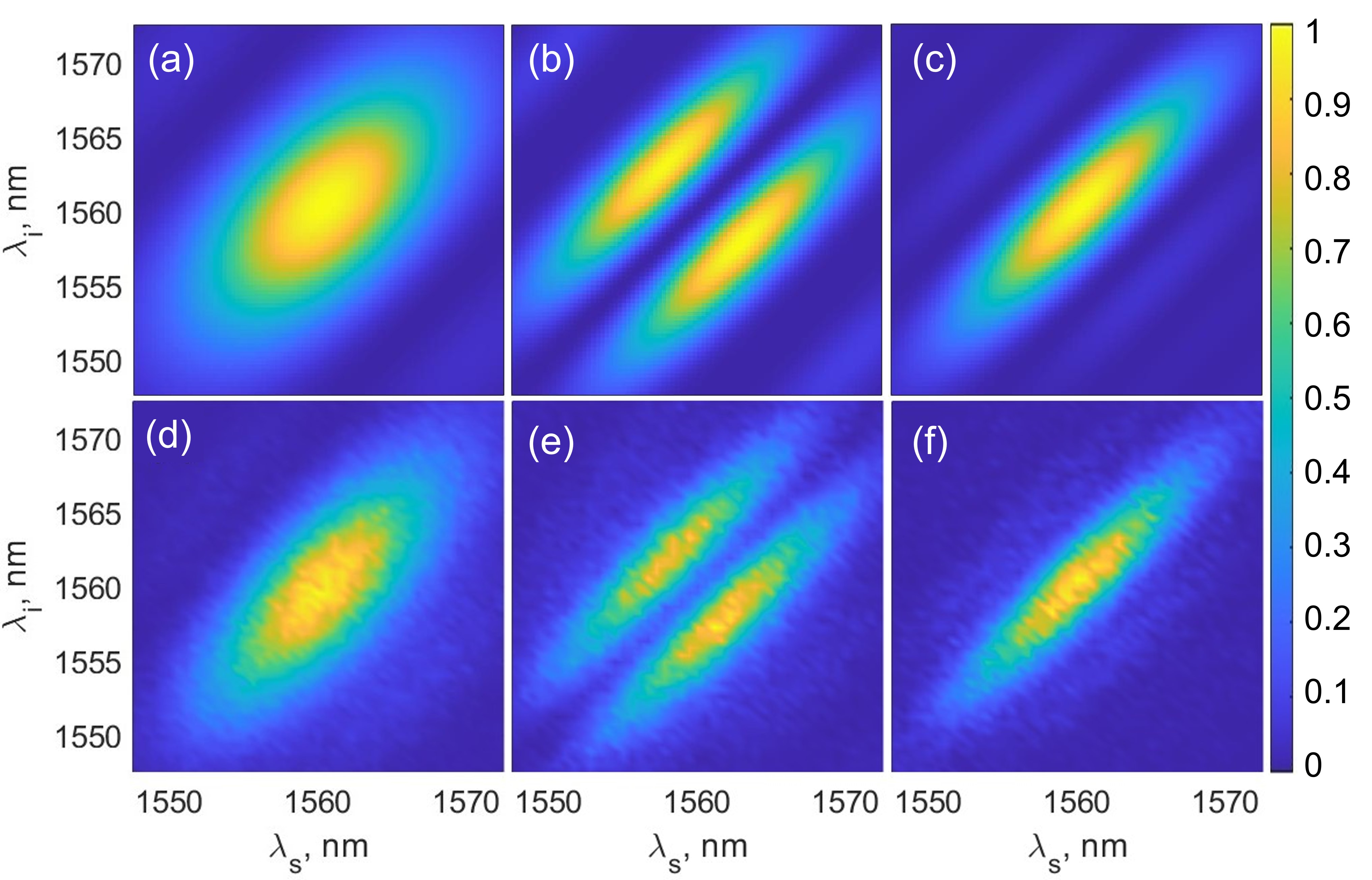}
         \label{fig:5mmHOMc}
\caption{Frequency-resolved Hong-Ou-Mandel interference from a 5\,mm long crystal. Top row - calculated JSIs, bottom row - experimentally measured JSIs. (a) and (d) The original JSI, measured without the HWP$_{3}$ and POL$_{2}$ in the setup  (see Fig.\,\ref{fig:setup}), (b) and (e) the JSI, measured by adding the HWP$_{3}$ at $22.5^{\circ}$ before the PBS, and (c) and (f) the JSI when the HWP$_{3}$ is replaced by the POL$_{2}$ at $45^{\circ}$. See Table\ref{fig:table} for crystal details}
\label{fig:5mmHOM}
\end{figure}

We demonstrated the above two methods using a 5\,mm long crystal with $\text{HG}_0$ PMF and a 15.9\,mm long crystal with $\text{HG}_1$ double-lobe PMF. The average walk-off delay values $\tau_{wo}$ of these crystals are $0.74~$ps (at 1560\,nm) and $2.36~$ps (at 1582\,nm), respectively.

Theoretically simulated and experimentally obtained results for the 5\,mm long crystal are shown at the top and bottom row in Fig.\,\ref{fig:5mmHOM}, respectively. After placing the HWP$_{3}$ at $22.5^{\circ}$, the original JSI (Fig.\,\ref{fig:5mmHOM}a,\,d), consisting of one lobe, splits into two lobes separated by the zero-intensity bunching (Fig.\,\ref{fig:5mmHOM}b,\,e), along the degenerate line. Measuring the JSI and its modulation frequency experimentally determines the temporal walk-off value or any other unknown delay between signal and idler photons.

The results when the HWP$_{3}$ is replaced by the POL$_{2}$ at $45^{\circ}$ are presented in Fig.\,\ref{fig:5mmHOM}c,\,f. As was derived above, in this case, anti-bunching is observed along the degenerate line.

Similar results for both methods were also obtained for a 15.9\,mm long $\text{HG}_0$ crystal (not presented). The longer crystal results in a proportionally longer walk-off delay, as well as in a narrower PMF's spectral width. As the periodicity of the spectral oscillations is inversely proportional to the walk-off delay, the total number of observed oscillations is preserved, regardless of the crystal length.

\begin{figure}
     \centering
         \includegraphics[width=0.7\columnwidth]{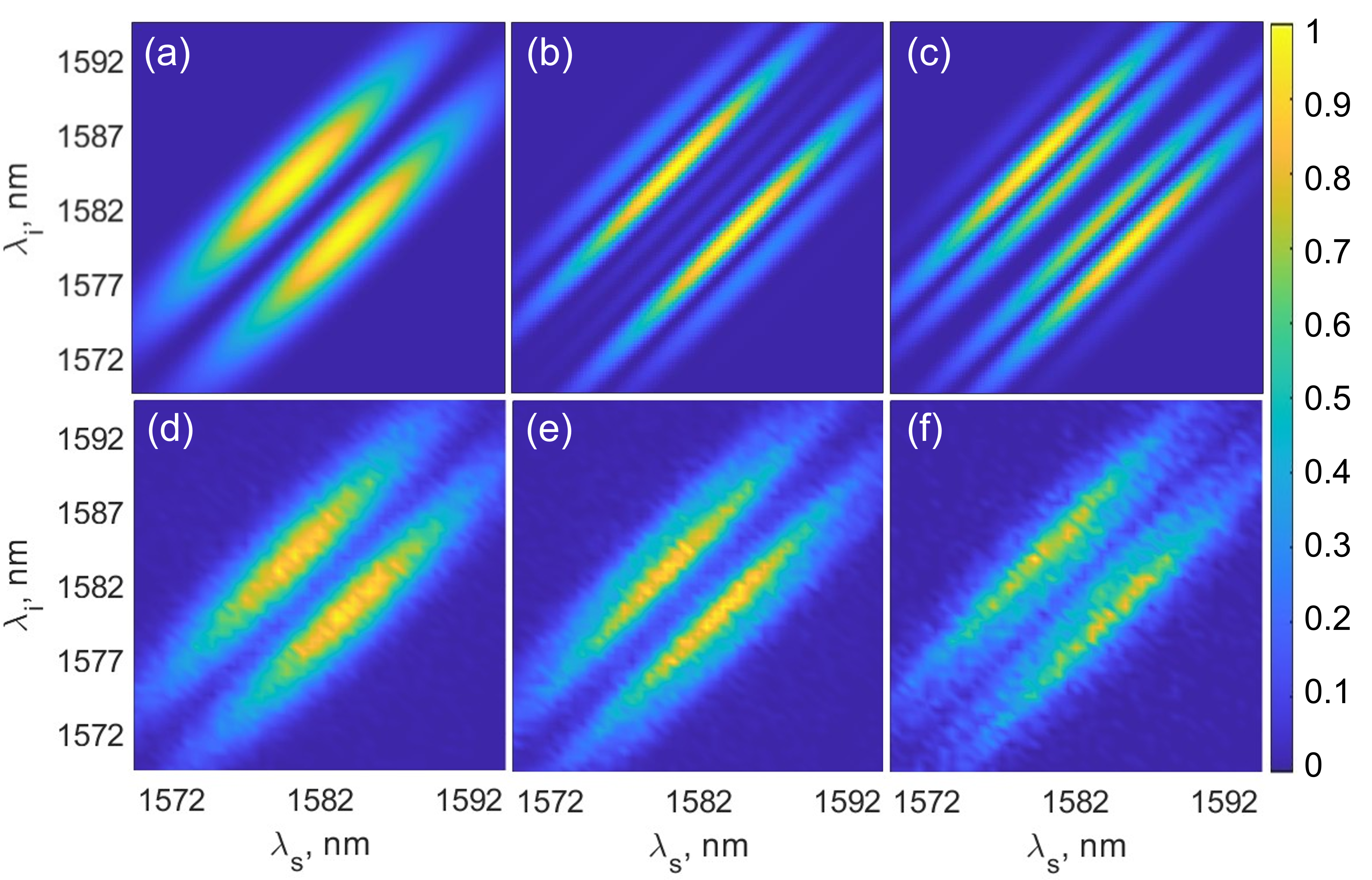}
\caption{Frequency-resolved Hong-Ou-Mandel interference from 15.9\,mm long crystal with the double-lobe PMF (see Table\ref{fig:table}, bottom line). Top row - calculated JSIs, bottom row - experimentally measured JSIs. (a) and (d) The original JSI, measured without the HWP$_{3}$ and POL$_{2}$ in the setup  (see Fig.\,\ref{fig:setup}), (b) and (e) the JSI, measured by adding the HWP$_{3}$ at $22.5^{\circ}$ before the PBS, and (c) and (f) the JSI when the HWP$_{3}$ is replaced by the POL$_{2}$ at $45^{\circ}$. The pump wavelength is 791 nm.}
\label{fig:15.9mmHOM}
\end{figure}

The JSI was measured for the other 15.9\,mm long crystal, which is characterized by the $\text{HG}_1$ PMF. Anti-symmetry of the PMF was experimentally revealed by Hong-Ou-Mandel interference. The JSIs, measured with the HWP$_{3}$ and POL$_{2}$ are the original JSI, modulated according to the symmetry of the PMF (see Eq.\,\ref{eq:JSI_HWP} and  Eq.\,\ref{eq:JSI_POL}). This unambiguously reveals the symmetry of the original JSA, which is anti-symmetric in our case. If the PMF was symmetric, the graphs for HWP and POL would be swapped.
The original JSI for this crystal is presented in Fig.\,\ref{fig:15.9mmHOM}a (simulated) and Fig.\,\ref{fig:15.9mmHOM}d (measured). As the walk-off is larger, but the PMF is as wide as for the shorter crystal, more oscillations are observed for the HWP and for the polarizer cases (Fig.\,\ref{fig:15.9mmHOM}b,c). Fitting the oscillations (using the walk-off value of 2.54 ps instead of the expected 2.36 ps due to the not perfectly collinear alignment of the setup or slightly different Sellmeier equations) for both measurements well agrees with the anti-symmetric $\text{HG}_1$ PMF. Thus, the presented methods can distinguish between symmetric and anti-symmetric PMF by bunching or anti-bunching the produced biphotons.

The walk-off $\tau_{wo}$ can be controlled by adding a birefringent crystal before the HWP or the polarizer. Different walk-off values lead to different modulation frequencies, thereby allowing one to further control the JSI along the difference-frequency axis in a similar way as pump modulation enables one to control the JSI along the sum-frequency axis.

\section{Conclusion}
We demonstrated type-II SPDC using both domain-engineered aperiodically poled nonlinear crystals, along with pump spectral modulation. This enables versatile control of the joint spectral intensity of photon pairs along the sum- and difference-frequency axes. Such control can generate novel quantum states in the spectral domain and stimulate the development of new protocols in quantum optics and quantum information science. 

Using those methods we have generated a variety of quantum states. Specifically, we have shown two of the four frequency-bin Bell states, but with minor modification the two additional Bell states can be generated. Moreover, one can attain two lobes like those shown in Fig.\,\ref{fig:15.9mm_2lobe}f but that do not overlap in wavelength, which increases the fidelity of the frequency-bin Bell state. Moreover, a dichroic mirror placed in the path of the photons described by such a JSI will create the near-degenerate polarization-entangled state $|\textrm{H}_{\omega_{1}}\textrm{V}_{\omega_{2}}\rangle-|\textrm{V}_{\omega_{1}}\textrm{H}_{\omega_{2}}\rangle$. Remarkably, a JSA consisting of non-overlapping lobes can be decomposed into separable Schmidt modes that can be easily filtered by regular spectral filters \cite{avella2014separable}. This results in a measurable Schmidt decomposition that enables one to tailor the degree of entanglement through Schmidt modes synthesis.

The 2D shaping methods of the JSI that we demonstrated here was used to generate states with interesting correlations between 4 pairs of signal and idler frequencies. Moreover, it can enable the creation of time-frequency GKP states, that are useful for quantum error correction \cite{fabre2020generation}.

Leveraging the Hong-Ou-Mandel interference enables further manipulation of the JSI along the difference-frequency axis after the photons' generation. Even if the photons do not overlap in time and time-domain Hong-Ou-Mandel experiments do not show any interference pattern, the two-photon interference still manifests in the multi-lobe JSI. This extends the dynamic range of interferometers and, along with a broadband pump, can be of a great advantage to quantum remote synchronization \cite{jin2016simple}, quantum-enhanced interferometric spectroscopy, metrology, and spectral domain quantum optical coherence tomography that are unattainable with classical states of light \cite{descamps2023quantum,kolenderska2020fourier,yepiz2020spectrally,kolenderska2020fourier,jin2016simple,chen2021temporal,dowling2008quantum,jin2018extended,li2023spectrally,triggiani2021quantum,chen2022quantum,rozema2014scalable}.

We also discussed that frequency-resolved Hong-Ou-Mandel interference can serve as a valuable tool for distinguishing between even and odd PMFs and measuring relative phases between the PMF's lobes. Experiments were done at the telecom wavelength of $1.5\,\mu$m, making the presented source compatible with optical fiber networks. When combined, our methods open a wide range of possibilities for preparing and manipulating quantum states of light for various tasks of quantum optics and quantum information technologies. The methods discussed in this paper can be easily generalized to crystals other than KTP and to nonlinear waveguides and various types of integrated photonic circuits where periodic poling is possible. Furthermore, with stronger pump power, the same setup can be used to shape bright squeezed vacuum \cite{hurvitz2023frequency,drago2022tunable}.

\begin{backmatter}
\bmsection{Funding}
This work is supported by the Israel Science Foundation (Grants No. 2085/18, 969/22), Israel Ministry of Science. 

\bmsection{Acknowledgments}
We thank Ziv Gefen from Raicol Crystals for assistance in preparing the nonlinear crystals.
I.H. acknowledges support by the Weinstein Research Institute for signal processing. I.H. acknowledges support by the Israeli planning and budgeting committee for quantum science and technology research.

\bmsection{Disclosures} The authors declare no conflicts of interest.

\bmsection{Data availability} The data that support the findings of this study are available from the corresponding author upon reasonable request.

\bigskip

\end{backmatter}


\bibliography{Optica-template}






\end{document}